\documentclass[onecolumn,11pt,preprintnumbers,amsmath,amssymb]{revtex4}
\usepackage{amsmath}
\usepackage{extarrows}
\usepackage[ruled]{algorithm2e}
\usepackage{graphicx}
\usepackage{dcolumn}
\usepackage{bm}
\usepackage{indentfirst}
\usepackage{amsmath}
\usepackage{multirow}
\usepackage{mathrsfs}
\usepackage{bbm}
\usepackage{euscript}
\usepackage{amssymb}
\usepackage{extarrows}
\usepackage{bbm}
\usepackage[ruled]{algorithm2e}
\usepackage{graphicx}
\usepackage{epstopdf}
\usepackage{dcolumn}
\usepackage{bm}
\usepackage{indentfirst}
\usepackage{amsmath}
\usepackage{multirow}
\usepackage{mathrsfs}
\usepackage{euscript}
\usepackage{amssymb}
\usepackage{appendix}
\usepackage{color,xcolor}

\begin{document}

\title{Photonic Energy-Coherence Theorem and Experimental Validations}

\author{Yan-Han Yang$^\dag$}
\author{Xin-Zhu Liu$^\dag$}
\author{Jun-Li Jiang}
\affiliation{Southwest Jiaotong University, School of Information Science and Technology, Chengdu, China, 610031}

\author{Hu Chen}

\affiliation{Southern University of Science and Technology, Shenzhen Institute for Quantum Science and Engineering, Shenzhen, China, 518055}

\author{Xue Yang$^*$}
\affiliation{Southwest Jiaotong University, School of Information Science and Technology, Chengdu, China, 610031}
\author{Xiao-Feng Qian$^*$}
\affiliation{Stevens Institute of Technology, Center for Quantum Science and Engineering and Department of Physics, Hoboken, New Jersey, USA, 07030}

\author{Shao-Ming Fei}
\affiliation{Capital Normal University, School of Mathematical Sciences, Beijing, China, 100048}
\affiliation{Max Planck Institute for Mathematics in the Sciences, Leipzig, Germany, 04103}

\author{Ming-Xing Luo$^*$}

\affiliation{Southwest Jiaotong University, School of Information Science and Technology, Chengdu, China, 610031}
\affiliation{Hefei National Laboratory,  University of Science and Technology of China,  Hefei, China, 230088}

\begin{abstract}
Wave-particle duality, intertwining two inherently contradictory properties of quantum systems, remains one of the most conceptually profound aspects of quantum mechanics. By using the concept of energy capacity, the ability of a quantum system to store and extract energy, we derive a device-independent uncertainty relation for wave-particle duality. This relation is shown to be independent of both the representation space and the measurement basis of the quantum system. Furthermore, we experimentally validate this wave-particle duality relation using a photon-based platform.   
\end{abstract}

\maketitle


{\noindent \footnotesize\textbf{*} Address all correspondence to Xue Yang, 
\href{mailto:yx12290552@swjtu.edu.cn}{yx12290552@swjtu.edu.cn};
Xiao-Feng Qian, \href{mailto:xqian6@stevens.edu}{xqian6@stevens.edu}; Ming-Xing Luo, 
\href{mailto:mxluo@swjtu.edu.cn}{mxluo@swjtu.edu.cn}
}

{\noindent \footnotesize\textbf{$\dagger$} These authors contributed equally to this work.}

\section{Introduction}
\label{sect:intro}  

Coherence, a fundamental property of waves, plays a crucial role in a wide range of applications, from electromagnetic and optical sciences \cite{BornWolf} to quantum information and quantum computing \cite{Plenio2017RMP}. Its significance was first demonstrated in Young's double-slit experiment, where light passing through two slits illuminated by the same source produced distinct interference patterns, confirming the wave nature of light\cite{BornWolf}. The quantum version of this experiment, using single photons, provided the first physical context for quantitative analysis of wave-particle duality \cite{Wootters-Zurek-79}. 

A self-consistent quantitative theory that captures both the wave and particle natures of quantum systems is essential for systematically characterizing their complementary nature. Numerous approaches have been explored in the literature. A primary focus has been on quantifying the ``waveness'' and ``particleness'' of a single quantum object and demonstrating their strong complementary relationship \cite{Greenberger-Yasin-88, Mandel-91,  Jaeger-Shimony-Vaidman-95, Englert-96}. These measures have been further shown to be connected to other coherence-related properties, such as entanglement \cite{Jacob-Bergou2010, QVE-18}, photon source purity \cite{Qian-Agarwal2020}, polarization coherence of light \cite{Luis-08, Eberly2017, FriebergPRL, Zela-18}, and even mechanical concepts of point masses \cite{Qian2023}. Another significant approach leverages Heisenberg's Uncertainty Principle \cite{Heisenberg1927}, which highlights the intrinsic uncertainty in simultaneously measuring two incompatible observables at the quantum level. This variance-based principle has been extended to measurement-independent relations using measurement probabilities \cite{Deutsch1983, Maassen1988, Friedland2013, Chen2018, UPReview, wu2022, xie2021}. Recent experimental advances have strengthened the connection between uncertainty relations and wave-particle duality \cite{spegel2024,verstraten2025}. However, most of these operational results depend on specific representation spaces or the measurement basis of the quantum system, limiting their general applicability.

\begin{figure}
\begin{center}
\begin{tabular}{c}
\includegraphics[height=4cm]{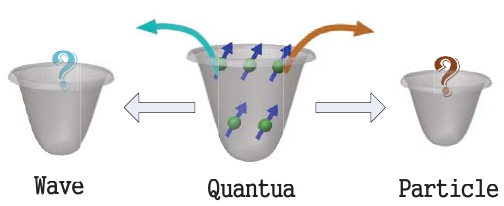}
\end{tabular}
\end{center}
\caption 
{ \label{fig1}
Schematically extracting energy of one photon in a delay-choice experiment. The goal is to characterize the maximal extractable energy of a given quantum state ($\rho$) in terms of its particle ($U_p\rho U_p^\dag$) and wave ($U_w\rho U_w^\dag$) configurations, defined under two unitary operations $U_p$ and $U_w$ with respect to any bare Hamiltonian. This approach establishes a device-independent wave-particle duality in terms of extractable energies, making the relation independent of both the representation space and the measurement basis of the quantum system. } 
\end{figure}

In this letter, we approach quantum duality from a distinctive perspective with quantities that are independent of both the representation space and the measurement basis. We propose a thermodynamics framework to characterize wave-particle duality by examining the maximal extractable energy of a quantum system during energy storage and supply processes, treating a single quantum particle as a quantum battery, as shown in Fig.~\ref{fig1}. This approach leverages the principles of quantum mechanics to enhance energy storage and conversion \cite{Allahverd2004,Binder2015,Perarnau2015,Andolina2019}. We demonstrate that the energy capacity \cite{Yang2023}, determined solely by the eigenvalues of the quantum state, establishes a clear physical connection to the intensities associated with both fringe visibility and the degree of distinguishability \cite{Eberly2017,Ionicioiu2011,Kaiser2012,Peruzzo2012,Ma2016}. We introduce a wave-particle duality as a squared energy coherence equality, which extends the conventional linear superposition of energy functionals \cite{Francica} and recent nonlinear relations \cite{Yang2023}. This provides the first characterization of the distinct roles played by wave and particle attributes in the functionality of quantum batteries, moving beyond the existing focus on quantum advantages in quantum batteries \cite{Binder2015, Ferraro2018, Seah2021, Francica2022, Rossini2020, Salvia2022}. To validate the theoretical findings, we provide experimental confirmation using single photons.

\section{Results}
\subsection{Energy wave-particle duality relations}
\label{sect:title}
Consider the well-known delay-choice experiment that consists of a polarization beam splitter (PBS), phase shifting (PS), beam merging (BM), and a photon source \cite{Eberly2017,Ionicioiu2011,Kaiser2012,Peruzzo2012,Ma2016}. The wave and particle features of a photonic state can be experimentally verified under the quantum control of a phase plate. Here, we investigate the experiment in terms of the extractable energies of a single photon, similar results can be extended for entangled setups \cite{Ma2016}. Specifically, a single photon emitted from a laser passes through some photonic devices and finally reaches the detectors. Suppose that the initial photon with polarization degree of freedom ($\mathrm{h}$ and $\mathrm{v}$ denote the horizontal and vertical polarizations respectively) is in the state given by
\begin{eqnarray}
\rho_{\rm{in}}=\sum_{i,j=\mathrm{h},\mathrm{v}}\rho_{ij}|i\rangle\langle j|=\frac{1}{2}\sum_{k=0}^3S_k\sigma_k,
\label{istatePauli}
\end{eqnarray}
where $S_i$'s are defined by $S_0=1$, $S_1=2|\rho_{12}|\cos \theta$, $S_2=2|\rho_{12}| \sin \theta$, and $S_3=\rho_{11}-\rho_{22}$, $\sigma_{0}$ denotes the identity matrix, and $\sigma_{k}$ are Pauli matrices. Take the governing Hamiltonian $H=E|\mathrm{h}\rangle\langle\mathrm{h}|$ as an example, where $E$ denotes the energy unit. The maximal extractable work capacity of a single photon is then given by (see details in the “Materials and methods” section): 
\begin{eqnarray}
\mathcal{C}_{\mathrm{p}}(\rho)=rE,
\label{workp}
\end{eqnarray}
where $r$ denotes the radius of the state on the Bloch sphere, i.e., $r=\sqrt{S_1^2+S_2^2+S_3^2}$. 

Now, consider the interferometric transformation $U_W$ that consists of three linear optical elements (PBS, PS, and BM) to explore the wave feature in the photonic delay-choice experiment \cite{Ionicioiu2011,Kaiser2012,Peruzzo2012,Ma2016}. Under the governing Hamiltonian $H$, the work capacity in terms of the wave features is given by (see details in the “Materials and methods” section): 
\begin{eqnarray}
\mathcal{C}_{\mathrm{v}}(\rho)=VE,
\label{workv}
\end{eqnarray}
where $V$ is the visibility \cite{Ionicioiu2011} defined by $V=\sqrt{S^2_2+S_3^2}$. 

Moreover, another unitary operation defined by  $U_{\pm}=(\sigma_1\pm \sigma_{3})/\sqrt{2}$ (one PBS) provides a two-way alternative to characterize the distinguishability of the particle features of single-photon \cite{Ionicioiu2011,Kaiser2012,Peruzzo2012,Ma2016}. This implies the work capacity in terms of the particle configuration (see details in the “Materials and methods” section) as:
\begin{eqnarray}
\mathcal{C}_{\mathrm{d}}(\rho)=|S_1|E.
\label{workd}
\end{eqnarray}
Combining with Eqs.~(\ref{workp},\ref{workv},\ref{workd}) implies a coherence uncertainty principle in terms of the extractable energy as 
\begin{eqnarray}
\max\{\mathcal{C}_{\mathrm{d}}(\rho),\mathcal{C}_{\mathrm{v}}(\rho)\}\leq \mathcal{C}_{\mathrm{p}}(\rho)\leq \mathcal{C}_{\mathrm{d}}(\rho)+\mathcal{C}_{\mathrm{v}}(\rho).
\label{worka}
\end{eqnarray}
Both qualities are not saturated for all $S_i$'s satisfying $S_i\not=0$. This implies that the work capacity of some quantum batteries cannot be determined according to only waves or particles. Interestingly, it further implies a strong energy duality equality as
\begin{eqnarray}
\mathcal{C}_{\mathrm{p}}^2(\rho)=\mathcal{C}_{\mathrm{d}}^2(\rho)+\mathcal{C}_{\mathrm{v}}^2(\rho).
\label{ECT}
\end{eqnarray}

In Young’s double-slit experiment, the brightness and darkness of the interference fringes can be quantified by the number of photons received per unit time. Eqs.~(\ref{workp}, \ref{workv}, \ref{workd}) offer similar interpretations of the delayed-choice experiment involving a single photon from the perspective of quantum thermodynamics. Specifically, the visibility of a single photon corresponds to the difference between the measured maximum and minimum quantum works in the wave configuration. Meanwhile, the distinguishability of the photon can be quantified by the maximum work difference in the particle configuration. This means Eqs.~(\ref{workp}, \ref{workv}, \ref{workd}) provide the first physically meaningful representations of the wave and particle characteristics in terms of quantum thermodynamics. 

The relation (\ref{ECT}) differs from the recent result \cite{Yang2023}, which was derived using various coherence measures in an operational manner. While our equality is obtained from polarized photons, it can be applied to any degree of freedom, such as spatial paths, energy levels, angular momentum states, or even atomic states. Additionally, we extend the present result to other bare Hamiltonian patterns, as detailed in the Supplementary Material. This establishes a device-independent wave-particle duality relation that stands apart from previous results \cite{Greenberger-Yasin-88, Mandel-91, Jaeger-Shimony-Vaidman-95, Englert-96, Qian-Agarwal2020, Luis-08, Eberly2017, FriebergPRL, Zela-18}, which typically depend on a specific representation space or the measurement basis of the quantum system.

It is worth noting that another linear coherence relation can be established using the recent method proposed in \cite{Francica}, which decomposes the total energy of a quantum battery into coherent and incoherent components. Specifically, this relation is expressed as $\mathcal{C}_{\mathrm{p}}(\rho) = \mathcal{C}_{\rm{coh}}(\rho) + \mathcal{C}_{\rm{incoh}}(\rho)$, where the work capacity $\mathcal{C}_{\rm{coh}}(\rho)$ is defined based on all unitary operations that preserve the coherence of the quantum state, and $\mathcal{C}_{\rm{incoh}}(\rho)$ represents the contribution from the incoherent part of the state \cite{Francica}. While this linear relation is reminiscent of classical systems, such as the decomposition of total energy into kinetic and potential energy, the equality (\ref{ECT}) presented here introduces a fundamentally different perspective. It highlights a quadratic superposition of the extractable energies of two-level quanta, expressed in terms of the wave and particle configurations in the standard quantum wave-particle experiment, offering deeper insights into the interplay between coherence and energy in quantum systems.

Additionally, this equality does not hold for the individual components of the maximal extractable energy or maximal injection energy \cite{Allahverd2004, Biswas2022, Tirone2022}, underscoring a key distinction between the present battery capacity and previous definitions based solely on maximal extractable energy \cite{Yang2023}. Moreover, it offers an alternative formulation of the Polarization Coherence Theorem \cite{Eberly2017}, reinterpreted in terms of the extractable work capacities (\ref{workp}, \ref{workv}, \ref{workd}) (see details in Supplementary Material).

\subsection{Experimental setup}

\begin{figure*}
\begin{center}
\begin{tabular}{c}
\includegraphics[height=4cm]{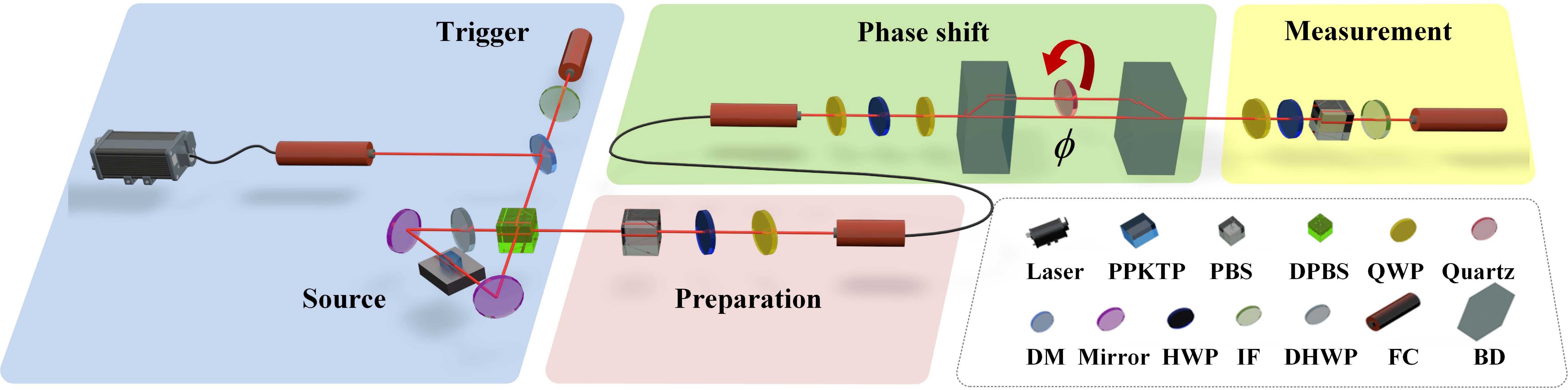}
\end{tabular}
\end{center}
\caption{\label{setup}
Experimental setup for verifying the relations in Eqs.~(\ref{worka}, \ref{ECT}). The first step is to prepare 810 nm two-photon entangled states generated by passing a 405 nm laser light through a type-II spontaneous parametric down-conversion and using a periodically poled $\rm{KTiOPO_4}$ (PPKTP) crystal (blue area). After the photons pass through a 3 nm interference filter (IF), one photon is designated as the trigger while the other signal photon is prepared in an arbitrary linear polarization state (pink area). Before the last part, the phase shift $\phi$ is adjusted by tilting the quartz plates between two beam displacers (BDs). The density matrix is constructed using quantum-state tomography (yellow area). The main elements consist of a half-wave plate (HWP), quarter-wave plate (QWP), polarizing beam splitter (PBS), dual-wavelength polarization beam splitter (DPBS), mirror (M), dichromatic mirror (DM), dual-wavelength half-wave plate (DHWP), and fiber coupler (FC).} 
\end{figure*}

We now propose a photon experiment to verify the present relations in Eqs.~(\ref{worka}, \ref{ECT}). The experimental setup, shown in Fig.~\ref{setup}, consists of four parts: the photon source, state preparation, phase shift, and measurement.

In the first step (blue area), a two-photon state is prepared where one photon acts as the trigger and the other as the signal photon. A cavity-stabilized Ti:sapphire pump laser, operating at a central wavelength of 405 nm with a spectral width of 0.03 nm, pumps a periodically poled $\rm{KTiOPO_4}$ (PPKTP) crystal to generate an 810 nm two-qubit photon pair via spontaneous parametric down-conversion (SPDC). The computational basis states $|0\rangle$ and $|1\rangle$ are encoded in horizontal ($|\mathrm{h}\rangle$) and vertical ($|\mathrm{v}\rangle$) polarizations, respectively. After passing through a dual-wavelength polarization beam splitter (DPBS), the pump beams are focused onto the center of the PPKTP crystal using a concave lens with a focal length of 20 cm. The crystal, with dimensions $15 * 2 * 1$ mm, operates under type-II phase matching, ensuring polarization entanglement between the photon pair. This photon source achieves high brightness (0.34 MHz) and a collection efficiency of $60\%$.

In the next step (pink area), a single-qubit arbitrary quantum state $|\phi\rangle = \alpha|\mathrm{h}\rangle + \beta|\mathrm{v}\rangle$ ($|\alpha|^2 + |\beta|^2 = 1$) is prepared. The signal photon passes through a half-wave plate (HWP) and a quarter-wave plate (QWP) to create the desired state. To correct polarization disturbances introduced by the single-mode fiber, the photon passes through a QWP-HWP-QWP sandwich structure, effectively mitigating these effects before interacting with further optical elements (green area). 

To obtain a phase shift (green area), an adjustable phase $\phi$ is introduced on the $|\mathrm{v}\rangle$ polarization by tilting quartz plates placed between two beam displacers (BDs). This results in the single-qubit state $|\phi\rangle = \alpha|\mathrm{h}\rangle + \beta e^{i\phi}|\mathrm{v}\rangle$. Finally, the photon is measured in the bases $\{|\mathrm{h}\rangle, |\mathrm{v}\rangle, |D\rangle =\frac{1}{\sqrt{2}} (|\mathrm{h}\rangle + |\mathrm{v}\rangle), |A\rangle = \frac{1}{\sqrt{2}}(|\mathrm{h}\rangle - |\mathrm{v}\rangle), |R\rangle = \frac{1}{\sqrt{2}}(|\mathrm{h}\rangle + i|\mathrm{v}\rangle), |L\rangle = \frac{1}{\sqrt{2}}(|\mathrm{h}\rangle - i|\mathrm{v}\rangle)\}$ (yellow area), by using a QWP, an HWP, and a polarization beam splitter (PBS).

\subsection{Statical analysis}

In the experiments, we prepare four initial quantum states:  
$|\phi_1\rangle = (0.5417 + 0.6645i)|\mathrm{h}\rangle + (-0.4545 + 0.2418i)|\mathrm{v}\rangle$,  
$|\phi_2\rangle = (0.6645 + 0.6797i)|\mathrm{h}\rangle + (-0.2418 - 0.1949i)|\mathrm{v}\rangle$,  
$|\phi_3\rangle = (0.6964 + 0.6124i)|\mathrm{h}\rangle + (-0.1228 + 0.3536i)|\mathrm{v}\rangle$, and  
$|\phi_4\rangle = (0.2418 + 0.5417i)|\mathrm{h}\rangle + (-0.6645 - 0.4545i)|\mathrm{v}\rangle$,  
by randomly setting the rotation angles of the HWP and QWP (pink area). Tomographic measurements are performed on the prepared single-photon states, with the post-selection based on 100 sets of photon coincidences. The measurement results are recorded in coincidence with the trigger photons. Using a 3 nm interference filter, the photon source generates up to 16,000 coincidence counts per second. The probability $p(a| x)$ of measurement outcomes $a \in \{0, 1\}$, conditional on the measurement setting $x$, is calculated as $p(a|x) = \frac{N_{x}^{a}}{N_{x}^{0} + N_{x}^{1}}$, where $N_{x}^{a}$ represents the photon coincidence counts. The density matrix is then reconstructed using the maximum-likelihood estimation \cite{ap2006,Jam2001}. All fidelities between the initial quantum states and ideal states exceed $98\%$, calculated according to the formula \cite{HHH}:  
\begin{eqnarray}
F(\rho_{\mathrm{exp}},\rho_{\mathrm{ideal}})= \mathrm{Tr}\left( \sqrt{\sqrt{\rho_{\mathrm{ideal}}} \rho_{\mathrm{exp}} \sqrt{\rho_{\mathrm{ideal}}}} \, \right) ^2
\end{eqnarray}
(see data in Supplementary Material). 

Note that the extractable energy of a single photon can be represented as $W = E \, \mathrm{Tr}(\rho |\mathrm{h}\rangle\langle \mathrm{h}|) \approx E p(a|x)$, where $E$ is the unit energy of each photon. This approximation is valid due to the stability of the single-photon source with a consistent wavelength. Thus, the extractable energy is experimentally estimated by using the photon coincidence counts except for the unit energy $E=2.45\times 10^{-19}$J of the involved photon. 

\begin{figure*}[!ht]
\begin{center}
\begin{tabular}{c}
\includegraphics[height=12.5cm]{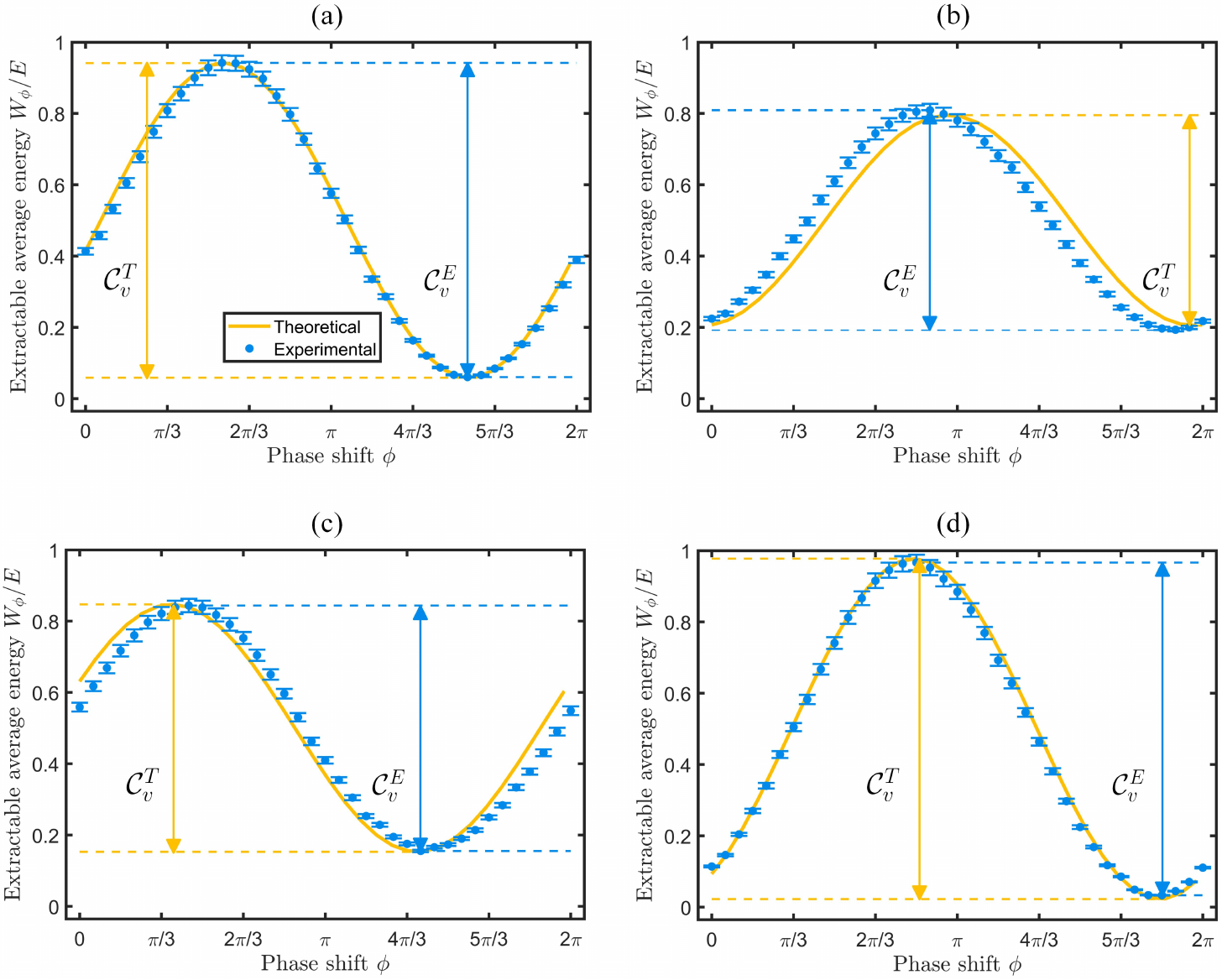}
\end{tabular}
\end{center}
\caption 
{ \label{FIGCV}
The extractable average energy $W_{\phi}$ for four initial states. (a) The initial state $|\phi_1\rangle$, (b) the initial state $|\phi_2\rangle$, (c) the initial state $|\phi_3\rangle$, and (d) the initial state $|\phi_4\rangle$. The blue points represent the experimental values of $W_{\phi}$, while the yellow lines denote the corresponding theoretical values. The difference between the maximal and minimal extractable average energies illustrates the work capacity $\mathcal{C}_{\mathrm{v}}$.} 
\end{figure*}

\subsection{Experimental results}
To determine the quantity  $\mathcal{C}_{\mathrm{v}}$, an additional phase shift $\phi$ is introduced to calculate the corresponding maximal and minimal extractable average energies, $W^{\rm v}_{\max}$ and $W^{\rm v}_{\min}$ (see definitions (\ref{Wv})). By tilting the quartz plates positioned between the two beam displacers (BDs) and scanning the phase $\phi$ over the interval $[0, 2\pi]$, the experimental maximal and minimal extractable average energies along $W_{\phi}$ curves for each prepared state are obtained as $(0.9420E,0.0605E), (0.8091E,0.1933E),(0.8439E,0.1553E)$ and $(0.9666E,0.0332E)$, as shown in Fig.~\ref{FIGCV}(a)-(d). The maximum discrepancy between the experimental and theoretical work capacity $\mathcal{C}_{\mathrm{v}}$ (from the difference between the maximal and minimal extractable average energies) is limited to 0.0253$E$ (blue points for experimental $W_{\phi}$, yellow lines for theoretical $W_{\phi}$ in Fig.~\ref{FIGCV}).

\begin{figure*}[htbp]
\begin{center}
\begin{tabular}{c}
\includegraphics[height=9cm]{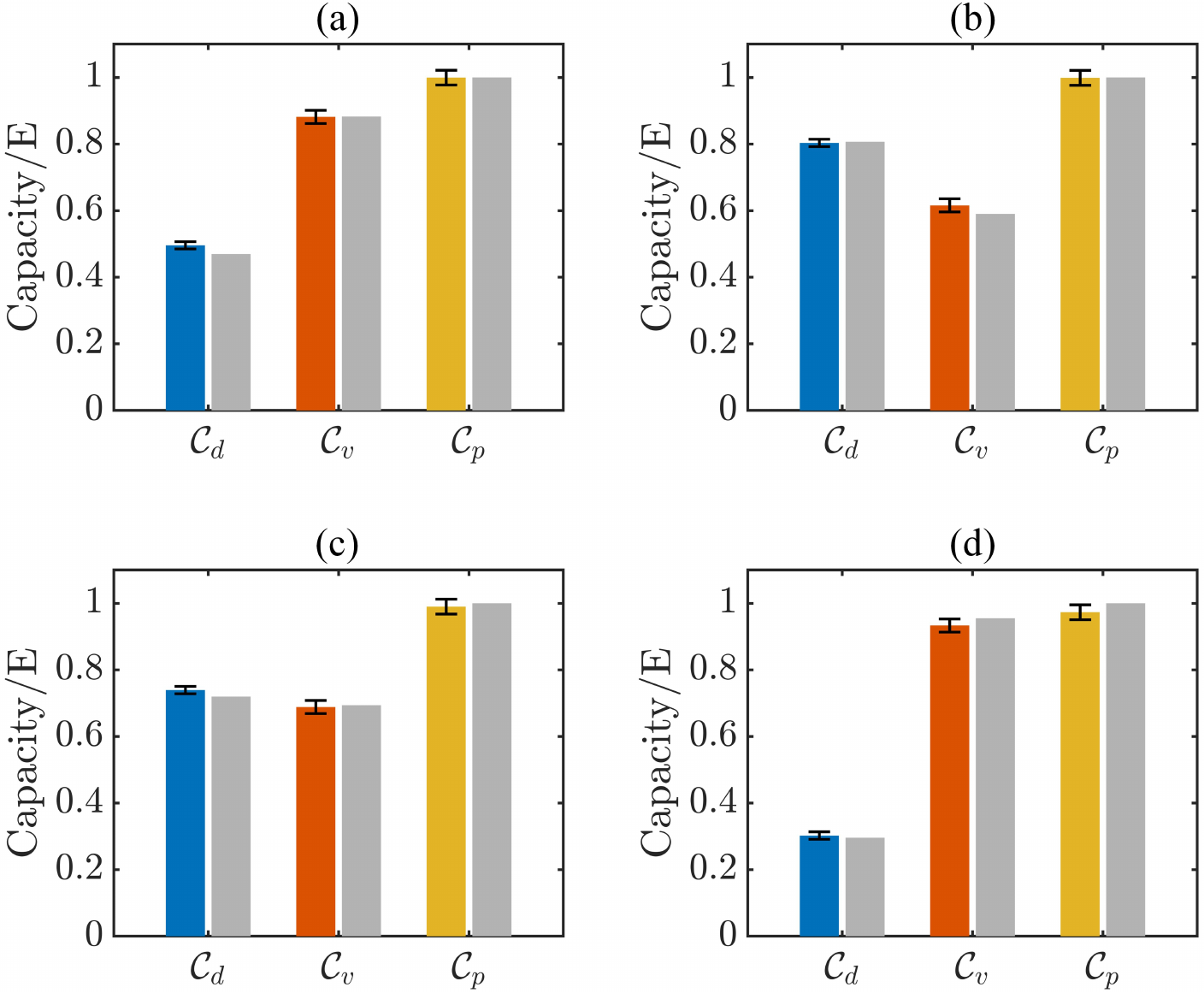}
\end{tabular}
\end{center}
\caption 
{ \label{FIGMM}
Verifying the work relation in Eq.~(\ref{worka}) with four initial quantum states (a) $|\phi_1\rangle=(0.5417+0.6645i)|\mathrm{h}\rangle+(-0.4545+0.2418i)|\mathrm{v}\rangle$, (b) $|\phi_2\rangle=(0.6645+0.6797i)|\mathrm{h}\rangle+(-0.2418-0.1949i)|\mathrm{v}\rangle$, (c) $|\phi_3\rangle=(0.6964+0.6124i)|\mathrm{h}\rangle+(-0.1228+0.3536i)|\mathrm{v}\rangle$, and (d)  $|\phi_4\rangle=(0.2418+0.5417i)|\mathrm{h}\rangle+(-0.6645-0.4545i)|\mathrm{v}\rangle$ with the bare Hamiltonian $H=E|\mathrm{h}\rangle\langle \mathrm{h}|$. In (a)-(d), experimental capacities are denoted by blue bars ($\mathcal{C}_{\mathrm{d}}/E$), red bars ($\mathcal{C}_{\mathrm{v}}/E$), yellow bars ($\mathcal{C}_{\mathrm{p}}/E$) and gray bars (corresponding theoretical values). The error bar is evaluated according to the experimental data.} 
\end{figure*}

Fig.~\ref{FIGMM}(a)-(d) demonstrate the relation (\ref{worka}) of four quantum states $|\phi_i\rangle$ using the bare Hamiltonian $H = E|\mathrm{h}\rangle \langle\mathrm{h}|$. The theoretical capacities $(\mathcal{C}_{\mathrm{d}},\mathcal{C}_{\mathrm{p}})$, calculated from the difference between the maximal and minimal extractable average energies, are given as $(0.4698E, E)$, $ (0.807E, E)$, $(0.7198E, E)$ and $(0.2962E, E)$ (gray bars). The experimental values of both $\mathcal{C}_{\mathrm{d}}$ and $\mathcal{C}_{\mathrm{p}}$ are evaluated by implementing Pauli measurements $\sigma_x$, $\sigma_y$, and $\sigma_z$ on each state. By tilting the quartz plates positioned between the two beam displacers (BDs) and scanning the phase $\phi$ over the interval $[0, 2\pi]$, all the capacities obtained along the curves $W_{\phi}$ for each state are given by $(0.4959E, 0.8816E, 0.9995E)$, $(0.8033E$, $0.6158E, 0.9988E)$, $(0.7395E, 0.6886E, 0.9902E)$ and $(0.3025E$, $0.9334E, 0.9731E)$ (blue bars for $\mathcal{C}_{\mathrm{d}}$, red bars for $\mathcal{C}_{\mathrm{v}}$, yellow bars for $\mathcal{C}_{\mathrm{p}}$), as shown in Fig.~\ref{FIGMM}. Both capacities $\mathcal{C}_{\mathrm{d}}$ and $\mathcal{C}_{\mathrm{v}}$ are smaller than $\mathcal{C}_{\mathrm{p}}$ while the combined capacity always exceeds $\mathcal{C}_{\mathrm{p}}$. This has verified the relation (\ref{worka}). The maximal discrepancy between the experimental and theoretical values is no more than 0.0269$E$.

\begin{figure*}[htbp]
\begin{center}
\begin{tabular}{c}
\includegraphics[height=6cm]{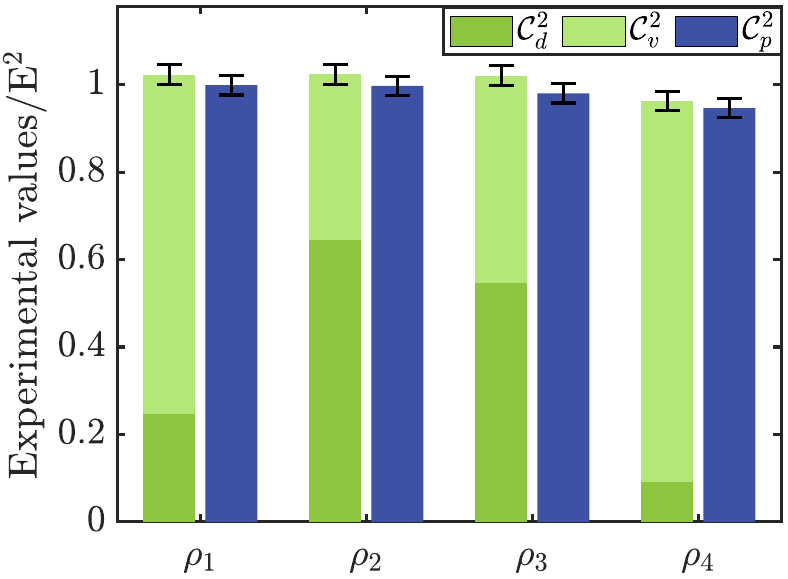}
\end{tabular}
\end{center}
\caption 
{ \label{FIGEQ}
The capacity equality (\ref{ECT}) for the initial quantum state $\rho_1=|\phi_1\rangle\langle\phi_1|$, $\rho_2=|\phi_2\rangle\langle\phi_2|$, $\rho_3=|\phi_3\rangle\langle\phi_3|$, and  $\rho_4=|\phi_4\rangle\langle\phi_4|$. Green bars represent experimental values of $\mathcal{C}_{\mathrm{d}}^2 +\mathcal{C}_{\mathrm{v}}^2$, while blue bars represent experimental values of $\mathcal{C}_{\mathrm{p}}^2$. The error bar is evaluated according to the experimental data.} 
\end{figure*}

For the same initial states, we finally verify the capacity equality (\ref{ECT}). The experimental values are shown in Fig.~\ref{FIGEQ}, where the experimental capacities are represented by green and blue bars, respectively. The present experiment can validate the equality (\ref{ECT}) for all initial states, with the error between the two sides caused by experimental statistics not exceeding 0.0406$E$.

\section{Discussion}
Rooted in quantum coherence, the quantum wave-particle duality has traditionally been quantified using specific measurements. The present duality relation does not rely on the specific representation space or measurement of the particle, offering a device-independent framework for wave-particle relations \cite{Plenio2017RMP}. This relation can also serve as a tool to witness the entanglement. While the wave-particle duality is often demonstrated with photons, our results extend to other physical particles \cite{Qian-Agarwal2020, Plenio2017RMP, Ma2016}, encouraging further experimental investigations with alternative platforms \cite{Le2018, Wenniger2022, Santos2019, Quach2022, Campaioli2017, Joshi2022, Hu2022, Gemme2024, Zhang2024}. Additionally, exploring the connection between extractable energies and other quantum features, such as entanglement and quantum advantages, remains an intriguing avenue for future research.

In this work, we proposed a quantitative approach to characterize wave-particle duality by examining the maximal extractable energy of a quantum system during both energy storage and supply processes. The resulting energy capacities establish a clear physical connection to fringe visibility and distinguishability. We introduced a novel wave-particle duality as a squared energy coherence equality, surpassing conventional linear relations of energy functionals. Finally, we proposed an experimental verification using single photons to explore the relationship between coherence and quantum batteries. These findings deepen our understanding of quantum mechanics and pave the way for advancements in quantum technologies and applications.

\section{Appendix: Materials and Methods}
\subsection{Energy Coherence Relations}
From the state (\ref{workp}),  both the minimal and maximal extractable average energies are given respectively by
\begin{eqnarray}
W_{\mathrm{p}}^{\min}&=&\frac{E}{2}\left[1-\sqrt{S_1^2+S_2^2+S_3^2}\right],
\nonumber\\
W_{\mathrm{p}}^{\max}&=&\frac{E}{2}\left[1+\sqrt{S_1^2+S_2^2+S_3^2}\right].
\label{workpmm}
\end{eqnarray}
This implies the work capacity as
\begin{eqnarray}
\mathcal{C}_{\mathrm{p}}=E\sqrt{S_1^2+S_2^2+S_3^2}.
\label{workpa}
\end{eqnarray}

Consider the most basic interferometric transformation $U_w$ that consists of three operations: polarization beam splitting (PBS), phase shifting (PS), and beam merging (BM). These operations can be respectively represented by the unitaries  $U_{\mathrm{bs}}=\frac{1}{\sqrt{2}}(\sigma_1+\sigma_{3})$ and $U_\phi=\exp(i\frac{\phi}{2}\sigma_3)$ 
for a shifting phase $\phi\in (0,2\pi)$. The unitary evolution $U_{w}:=U_{\mathrm{bs}} U_{\phi}U_{\mathrm{bs}}^\dag$ will change the incoming photon into the following state:
\begin{eqnarray}
\rho_{\mathrm{out}}=\frac{1}{2}\sum_{k=0}^3\hat{S}_k\sigma_k,
\label{rhoout}
\end{eqnarray}
where $\hat{S}_k$ are given by $\hat{S}_0=1$, $\hat{S}_1=S_1$, $\hat{S}_2=S_2 \cos\phi +S_3 \sin\phi$, and $\hat{S}_3=-S_2 \sin\phi+S_3 \cos\phi$. Suppose that the bare Hamiltonian is given $H=E|\mathrm{h}\rangle\langle\mathrm{h}|$, i.e., the upper lag of the detector with the eigenstate $|\mathrm{h}\rangle\langle \mathrm{h}|$ of  $\sigma_3$. The extractable energy from the output state is shown as
\begin{eqnarray}
W_{\phi}&=&{\rm Tr}(H\rho_{\mathrm{out}})
\nonumber
\\
&=&\frac{E}{2}(1-S_2\sin\phi+S_3\cos\phi)
\nonumber
\\
&=&
\frac{E}{2}(1+V \cos(\phi-\varphi)),
\label{Wphi}
\end{eqnarray}
where $\varphi$ is defined according to $\varphi=\tan^{-1}(S_3/S_2)$,  and $V=\sqrt{S^2_2+S_3^2}$. This implies that the maximal and minimal extractable average energies as
\begin{eqnarray}
W^{\mathrm{v}}_{\max}&=&\max_{\phi}W_{\phi}=\frac{E}{2}(1+V),
\nonumber\\
W^{\mathrm{v}}_{\min}&=&\min_{\phi}W_{\phi}=\frac{E}{2}(1-V).
\label{Wv}
\end{eqnarray}
This yields to the work capacity as 
\begin{eqnarray}
\mathcal{C}_{\mathrm{v}}=W^{\mathrm{v}}_{\max}-W^{\mathrm{v}}_{\min}=EV.
\label{workva}
\end{eqnarray}

As for the particle distinguishability, consider another unitary operation $U_{\mathrm{d}}\in \{\frac{1}{\sqrt{2}}(\sigma_1\pm \sigma_{3})\}$ of one PBS. We obtain the work capacity as
\begin{eqnarray}
\mathcal{C}_{\mathrm{d}}=W^{\mathrm{d}}_{\max}-W^{\mathrm{d}}_{\min}=E|S_1|,
\label{workda}
\end{eqnarray}
where the maximal and minimal extractable energies are respectively given by $\frac{1}{2}(1+|S_1|)$ and $\frac{1}{2}(1-|S_1|)$.

\subsection{Device-independent wave-particle duality}

As the proof in the ``Work Capacity Theorem'' section is independent of the quantum state, it is sufficient to show that the present duality is measurement-independent. Specifically, consider the bare Hamiltonian as $H=E|0\rangle_1\langle 0|$ with $|0\rangle_1$ being the up eigenvector of $\sigma_1$, namely, $|0\rangle_1=\frac{1}{\sqrt{2}}(|0\rangle_3+|1\rangle_3)$, we use the notation $\sigma_i|j\rangle=(-1)^j|j\rangle_i$ ($i=1, 2, 3$; $j=0, 1$) for the eigenvectors of the Pauli matrices. Implement a new phase shifting by $U_\phi=\exp(i \frac{\phi}{2}\sigma_3)$. Equation (\ref{rhoout}) still holds but now with $\hat{S}_0=1$, $\hat{S}_1=S_1 \cos\phi + S_2 \sin \phi$, $\hat{S}_2=-S_1 \sin \phi+ S_2 \cos\phi$, and $\hat{S}_3=S_3$. The average energy from the output state is then given as
\begin{eqnarray}
\hat{W}_{\phi}&=&{\rm Tr}(H\rho_{out})
\nonumber\\
&=&\frac{E}{2}\left(
1+S_1\sin\phi+S_2\cos\phi
\right)
\nonumber\\
&=&
\frac{E}{2}(1+V\cos(\phi-\varphi)),
\label{eqn25}
\end{eqnarray}
where $\varphi$ is defined by $\varphi=\tan^{-1}(S_2/S_1)$ and $V=\sqrt{S^2_1+S_2^2}$. This implies the maximal and minimal energies given by
\begin{eqnarray}
&&W^{\mathrm{v}}_{\max}=\max_{\phi}\hat{W}_{\phi}=\frac{E}{2}(1+V),
\nonumber\\
&&W^{\mathrm{v}}_{\min}=\min_{\phi}\hat{W}_{\phi}=\frac{E}{2}(1-V).
\label{eqn26}
\end{eqnarray}
We then get the work capacity as
\begin{eqnarray}
\mathcal{C}_{\mathrm{v}}=W^{\mathrm{v}}_{\max}-W^{\mathrm{v}}_{\min}=VE.
\label{eqn27}
\end{eqnarray}

As for the distinguishability of the particle configuration, with the unitary operations $U_{\mathrm{d}}\in \{\frac{1}{\sqrt{2}}(\sigma_1\pm \sigma_{3})\}$ the maximal and minimal energies can be extracted from the final state as 
\begin{eqnarray}
&&W^{\mathrm{d}}_{\max}=\max_{U_{\mathrm{p}}\in \{U_{\pm}\}}{\rm Tr}(H U_{\mathrm{p}}\rho_{\rm in}U_{\mathrm{p}}^\dag)=\frac{E}{2}(1+|S_3|),
\nonumber\\
&&W^{\mathrm{d}}_{\min}=\min_{U_d\in \{U_{\pm}\}}{\rm Tr}(H U_d\rho_{\rm in}U_d^\dag)=\frac{E}{2}(1-|S_3|).
\label{eqn28}
\end{eqnarray}
This follows the energy capacity as
\begin{eqnarray}
\mathcal{C}_{\mathrm{d}}=W^{\mathrm{d}}_{\max}-W^{\mathrm{d}}_{\min}=|S_3|E.
\label{eqn29}
\end{eqnarray} 
This implies a similar wave-particle duality of $\mathcal{C}_{\mathrm{p}}^2=\mathcal{C}_{\mathrm{d}}^2+\mathcal{C}_{\mathrm{v}}^2$ even if all the quantities of $\mathcal{C}_{\mathrm{p}}$, $\mathcal{C}_{\mathrm{d}}$ and $\mathcal{C}_{\mathrm{v}}$ are different from these obtained by using the bare Hamiltonian $\hat{H}=E|\mathrm{h}\rangle\langle \mathrm{h}|$ . 

Moreover, for a given bare Hamiltonian $\hat{H}=E|\varphi\rangle\langle \varphi|$ with the qubit state $|\varphi\rangle$, we obtain the equality 
of ${\rm Tr}(\rho \hat{H})={\rm Tr}(U\rho U^\dag  H)$, where $U$ denotes the unitary transformation satisfying $U: |\varphi\rangle\mapsto |0\rangle_1$. This combined with the source independence implies a similar wave-particle duality of 
\begin{eqnarray}
\mathcal{C}_{\mathrm{p}}^2=\mathcal{C}_{\mathrm{d}}^2+\mathcal{C}_{\mathrm{v}}^2
\end{eqnarray}
for any bare Hamiltonian $\hat{H}$.

\subsection*{Disclosures}
The authors declare no competing financial interests.

\acknowledgments 
We thank the discussions with Prof. Yunlong Xiao, Zhihao Ma, and Zhengda Li. This work was supported by the National Natural Science Foundation of China (Nos.62172341, 12405024, 12204386, and 12171044), National Natural Science Foundation of Sichuan Provence (Nos. 2024NSFSC1365,2024NSFSC1375), Innovation Program for Quantum Science and Technology (No. 2021ZD0301601), the Academician Innovation Platform of Hainan Province. XQ acknowledges support from NSF Grant No. PHY-2316878.



\begin{thebibliography}{10}

\bibitem{BornWolf}
M.~Born and E.~Wolf, {\em Principles of optics: electromagnetic theory of propagation, interference and diffraction of light}, Elsevier  (2013).

\bibitem{Plenio2017RMP}
A.~Streltsov, G.~Adesso, and M.~B. Plenio, ``Colloquium: Quantum coherence as a resource,'' {\em Reviews of Modern Physics} {\bf 89}(4), 041003  (2017).

\bibitem{Wootters-Zurek-79}
W.~K. Wootters and W.~H. Zurek, ``Complementarity in the double-slit experiment: Quantum nonseparability and a quantitative statement of bohr's principle,'' {\em Physical Review D} {\bf 19}(2), 473  (1979).

\bibitem{Greenberger-Yasin-88}
D.~M. Greenberger and A.~Yasin, ``Simultaneous wave and particle knowledge in a neutron interferometer,'' {\em Physics Letters A} {\bf 128}(8), 391--394  (1988).

\bibitem{Mandel-91}
L.~Mandel, ``Coherence and indistinguishability,'' {\em Optics Letters} {\bf 16}(23), 1882--1883  (1991).

\bibitem{Jaeger-Shimony-Vaidman-95}
G.~Jaeger, A.~Shimony, and L.~Vaidman, ``Two interferometric complementarities,'' {\em Physical Review A} {\bf 51}(1), 54  (1995).

\bibitem{Englert-96}
B.-G. Englert, ``Fringe visibility and which-way information: An inequality,'' {\em Physical Review Letters} {\bf 77}(11), 2154  (1996).

\bibitem{Jacob-Bergou2010}
M.~Jakob and J.~A. Bergou, ``Quantitative complementarity relations in bipartite systems: Entanglement as a physical reality,'' {\em Optics Communications} {\bf 283}(5), 827--830  (2010).

\bibitem{QVE-18}
X.-F. Qian, A.~Vamivakas, and J.~Eberly, ``Entanglement limits duality and vice versa,'' {\em Optica} {\bf 5}(8), 942--947  (2018).

\bibitem{Qian-Agarwal2020}
X.-F. Qian and G.~Agarwal, ``Quantum duality: A source point of view,'' {\em Physical Review Research} {\bf 2}(1), 012031  (2020).

\bibitem{Luis-08}
A.~Luis, ``Quantum-classical correspondence for visibility, coherence, and relative phase for multidimensional systems,'' {\em Physical Review A} {\bf 78}(2), 025802  (2008).

\bibitem{Eberly2017}
J.~Eberly, X.-F. Qian, and A.~Vamivakas, ``Polarization coherence theorem,'' {\em Optica} {\bf 4}(9), 1113--1114  (2017).

\bibitem{FriebergPRL}
A.~Norrman, K.~Blomstedt, T.~Set{\"a}l{\"a}, {\em et~al.}, ``Complementarity and polarization modulation in photon interference,'' {\em Physical Review Letters} {\bf 119}(4), 040401  (2017).

\bibitem{Zela-18}
F.~De~Zela, ``Hidden coherences and two-state systems,'' {\em Optica} {\bf 5}(3), 243--250  (2018).

\bibitem{Qian2023}
X.-F. Qian and M.~Izadi, ``Bridging coherence optics and classical mechanics: A generic light polarization-entanglement complementary relation,'' {\em Physical Review Research} {\bf 5}(3), 033110  (2023).

\bibitem{Heisenberg1927}
W.~Heisenberg, ``{\"U}ber den anschaulichen inhalt der quantentheoretischen kinematik und mechanik,'' {\em Zeitschrift f{\"u}r Physik} {\bf 43}(3), 172--198  (1927).

\bibitem{Deutsch1983}
D.~Deutsch, ``Uncertainty in quantum measurements,'' {\em Physical Review Letters} {\bf 50}(9), 631  (1983).

\bibitem{Maassen1988}
H.~Maassen and J.~B. Uffink, ``Generalized entropic uncertainty relations,'' {\em Physical Review Letters} {\bf 60}(12), 1103  (1988).

\bibitem{Friedland2013}
S.~Friedland, V.~Gheorghiu, and G.~Gour, ``Universal uncertainty relations,'' {\em Physical review letters} {\bf 111}(23), 230401  (2013).

\bibitem{Chen2018}
Z.~Chen, Z.~Ma, Y.~Xiao, {\em et~al.}, ``Improved quantum entropic uncertainty relations,'' {\em Physical Review A} {\bf 98}(4), 042305  (2018).

\bibitem{UPReview}
P.~J. Coles, M.~Berta, M.~Tomamichel, {\em et~al.}, ``Entropic uncertainty relations and their applications,'' {\em Reviews of Modern Physics} {\bf 89}(1), 015002  (2017).

\bibitem{wu2022}
L.~Wu, L.~Ye, and D.~Wang, ``Tighter generalized entropic uncertainty relations in multipartite systems,'' {\em Physical Review A} {\bf 106}(6), 062219  (2022).

\bibitem{xie2021}
B.-F. Xie, F.~Ming, D.~Wang, {\em et~al.}, ``Optimized entropic uncertainty relations for multiple measurements,'' {\em Physical Review A} {\bf 104}(6), 062204  (2021).

\bibitem{spegel2024}
D.~Spegel-Lexne, S.~G{\'o}mez, J.~Argillander, {\em et~al.}, ``Experimental demonstration of the equivalence of entropic uncertainty with wave-particle duality,'' {\em Science advances} {\bf 10}(49), eadr2007  (2024).

\bibitem{verstraten2025}
J.~Verstraten, K.~Dai, M.~Dixmerias, {\em et~al.}, ``In situ imaging of a single-atom wave packet in continuous space,'' {\em Physical Review Letters} {\bf 134}(8), 083403  (2025).

\bibitem{Allahverd2004}
A.~E. Allahverdyan, R.~Balian, and T.~M. Nieuwenhuizen, ``Maximal work extraction from finite quantum systems,'' {\em Europhysics Letters} {\bf 67}(4), 565  (2004).

\bibitem{Binder2015}
F.~C. Binder, S.~Vinjanampathy, K.~Modi, {\em et~al.}, ``Quantacell: powerful charging of quantum batteries,'' {\em New Journal of Physics} {\bf 17}(7), 075015  (2015).

\bibitem{Perarnau2015}
M.~Perarnau-Llobet, K.~V. Hovhannisyan, M.~Huber, {\em et~al.}, ``Extractable work from correlations,'' {\em Physical Review X} {\bf 5}(4), 041011  (2015).

\bibitem{Andolina2019}
G.~M. Andolina, M.~Keck, A.~Mari, {\em et~al.}, ``Extractable work, the role of correlations, and asymptotic freedom in quantum batteries,'' {\em Physical Review Letters} {\bf 122}(4), 047702  (2019).

\bibitem{Yang2023}
X.~Yang, Y.-H. Yang, M.~Alimuddin, {\em et~al.}, ``Battery capacity of energy-storing quantum systems,'' {\em Physical Review Letters} {\bf 131}(3), 030402  (2023).

\bibitem{Ionicioiu2011}
R.~Ionicioiu and D.~R. Terno, ``Proposal for a quantum delayed-choice experiment,'' {\em Physical Review Letters} {\bf 107}(23), 230406  (2011).

\bibitem{Kaiser2012}
F.~Kaiser, T.~Coudreau, P.~Milman, {\em et~al.}, ``Entanglement-enabled delayed-choice experiment,'' {\em Science} {\bf 338}(6107), 637--640  (2012).

\bibitem{Peruzzo2012}
A.~Peruzzo, P.~Shadbolt, N.~Brunner, {\em et~al.}, ``A quantum delayed-choice experiment,'' {\em Science} {\bf 338}(6107), 634--637  (2012).

\bibitem{Ma2016}
X.-s. Ma, J.~Kofler, and A.~Zeilinger, ``Delayed-choice gedanken experiments and their realizations,'' {\em Reviews of Modern Physics} {\bf 88}(1), 015005  (2016).

\bibitem{Francica}
G.~Francica, F.~C. Binder, G.~Guarnieri, {\em et~al.}, ``Quantum coherence and ergotropy,'' {\em Physical Review Letters} {\bf 125}(18), 180603  (2020).

\bibitem{Ferraro2018}
D.~Ferraro, M.~Campisi, G.~M. Andolina, {\em et~al.}, ``High-power collective charging of a solid-state quantum battery,'' {\em Physical Review Letters} {\bf 120}(11), 117702  (2018).

\bibitem{Seah2021}
S.~Seah, M.~Perarnau-Llobet, G.~Haack, {\em et~al.}, ``Quantum speed-up in collisional battery charging,'' {\em Physical Review Letters} {\bf 127}(10), 100601  (2021).

\bibitem{Francica2022}
G.~Francica, ``Quantum correlations and ergotropy,'' {\em Physical Review E} {\bf 105}(5), L052101  (2022).

\bibitem{Rossini2020}
D.~Rossini, G.~M. Andolina, D.~Rosa, {\em et~al.}, ``Quantum advantage in the charging process of sachdev-ye-kitaev batteries,'' {\em Physical Review Letters} {\bf 125}(23), 236402  (2020).

\bibitem{Salvia2022}
R.~Salvia, M.~Perarnau-Llobet, G.~Haack, {\em et~al.}, ``Quantum advantage in charging cavity and spin batteries by repeated interactions,'' {\em Physical Review Research} {\bf 5}(1), 013155  (2023).

\bibitem{Biswas2022}
T.~Biswas, M.~{\L}obejko, P.~Mazurek, {\em et~al.}, ``Extraction of ergotropy: Free energy bound and application to open cycle engines,'' {\em Quantum} {\bf 6}, 841  (2022).

\bibitem{Tirone2022}
S.~Tirone, R.~Salvia, S.~Chessa, {\em et~al.}, ``Quantum work capacitances,'' {\em arXiv preprint arXiv:2211.02685}   (2022).

\bibitem{ap2006}
J.~B. Altepeter, E.~R. Jeffrey, and P.~G. Kwiat, ``Photonic state tomography,'' {\em Advances in Atomic, Molecular, and Optical Physics} {\bf 52}, 105--159  (2005).

\bibitem{Jam2001}
D.~F. James, P.~G. Kwiat, W.~J. Munro, {\em et~al.}, ``Measurement of qubits,'' {\em Physical Review A} {\bf 64}(5), 052312  (2001).

\bibitem{HHH}
R.~Horodecki, P.~Horodecki, M.~Horodecki, {\em et~al.}, ``Quantum entanglement,'' {\em Reviews of Modern Physics} {\bf 81}(2), 865--942  (2009).

\bibitem{Le2018}
T.~P. Le, J.~Levinsen, K.~Modi, {\em et~al.}, ``Spin-chain model of a many-body quantum battery,'' {\em Physical Review A} {\bf 97}(2), 022106  (2018).

\bibitem{Wenniger2022}
I.~M. de~Buy~Wenniger, S.~Thomas, M.~Maffei, {\em et~al.}, ``Coherence-powered work exchanges between a solid-state qubit and light fields,'' {\em Physical Review Letters} {\bf 131}(26), 260401  (2023).

\bibitem{Santos2019}
A.~C. Santos, B.~{\c{C}}akmak, S.~Campbell, {\em et~al.}, ``Stable adiabatic quantum batteries,'' {\em Physical Review E} {\bf 100}(3), 032107  (2019).

\bibitem{Quach2022}
J.~Q. Quach, K.~E. McGhee, L.~Ganzer, {\em et~al.}, ``Superabsorption in an organic microcavity: Toward a quantum battery,'' {\em Science Advances} {\bf 8}(2), eabk3160  (2022).

\bibitem{Campaioli2017}
F.~Campaioli, F.~A. Pollock, F.~C. Binder, {\em et~al.}, ``Enhancing the charging power of quantum batteries,'' {\em Physical review letters} {\bf 118}(15), 150601  (2017).

\bibitem{Joshi2022}
J.~Joshi and T.~Mahesh, ``Experimental investigation of a quantum battery using star-topology nmr spin systems,'' {\em Physical Review A} {\bf 106}(4), 042601  (2022).

\bibitem{Hu2022}
C.-K. Hu, J.~Qiu, P.~J. Souza, {\em et~al.}, ``Optimal charging of a superconducting quantum battery,'' {\em Quantum Science and Technology} {\bf 7}(4), 045018  (2022).

\bibitem{Gemme2024}
G.~Gemme, M.~Grossi, S.~Vallecorsa, {\em et~al.}, ``Qutrit quantum battery: Comparing different charging protocols,'' {\em Physical Review Research} {\bf 6}(2), 023091  (2024).

\bibitem{Zhang2024}
J.-W. Zhang, B.~Wang, W.-F. Yuan, {\em et~al.}, ``Energy-conversion device using a quantum engine with the work medium of two-atom entanglement,'' {\em Physical Review Letters} {\bf 132}(18), 180401  (2024).

\end{thebibliography}
\end{document}